\documentclass[12pt]{iopart}

\begin{document}

\title[Comment on ``Small Lorentz violations...''  ]{Comment on ``Small Lorentz violations in quantum gravity: do they lead to unacceptably large effects?''}

\author{Joseph Polchinski}

\address{KITP, University of California, Santa Barbara, CA 93106-4030, USA}
\ead{joep@kitp.ucsb.edu}
\begin{abstract}
A recent paper by Gambini, Rastgoo and Pullin~\cite{grp} investigates the important issue of constraints from Lorentz invariance on Planck scale physics, arguing that the classic analysis of Collins, Perez, Sudarsky, Urrutia and Vucetich~\cite{cpsuv} is not generally valid.  We argue that the new work is based on models that do not capture the relevant physics, and that almost all models of observable high energy Lorentz violation, and proposed Lorentz-violating theories of quantum gravity, are ruled out by low energy tests; the only known exceptions are based on supersymmetry.

\end{abstract}

\maketitle

The high precision with which Lorentz invariance is observed in nature places strong constraints on what can happen at much higher energies.  It is a general principle of local quantum theory that physics at a high scale $M$ manifests itself at lower energies through effective local terms in the action.  A local operator of dimension $\Delta$ is induced with a coefficient of order $M^{4 - \Delta}$, the 4 being the dimension of spacetime.  Generically all operators allowed by symmetry are generated.  Thus, operators of dimension $\leq 4$ provide a direct window onto the symmetries of the high energy theory, no matter how high the scale of symmetry breaking.

The Standard Model admits a large number of dimension 4 operators that are gauge invariant but not Lorentz invariant, for example the spatial gradient terms for each of the 19 gauge multiplets.  These lead to different `speeds of light' for the different multiplets, so that Lorentz breaking of order one at high energy would lead to unacceptably large breaking at low energy.  This reasoning has been confirmed in a model calculation in Ref.~\cite{cpsuv} (hereafter denoted CPSUV), which also reviews related work.  The effective symmetry-breaking interactions are suppressed by Standard Model coupling factors, but are still far too large.  Heuristically one can think of a low energy particle mixing with a highly virtual pair with momenta near $M$: the symmetry breaking at the high scale feeds down to low energy without suppression due to the dimensional argument above.

Ref.~\cite{grp} (hereafter denoted GRP) reexamines the issue raised by CPSUV, finding in two models that the Lorentz violation is small, and concluding that this result will hold rather generally.  The purpose of this comment is to note that the two models considered have special features that are not present in the theories of interest.  Indeed, without these the special features the calculations of GRP support the claims of CPSUV.

The first model is a Euclidean lattice theory, where the ratios of the time and space lattice steps are taken equal in the continuum limit.\footnote{The issues discussed in this paragraph were already raised in CPSUV.}  A Euclidean lattice with equal steps along different axes has discrete rotational symmetries, which forbids the dimension 4 terms that would violate the Euclidean Lorentz ({\it i.e.}\ rotational) invariance; indeed, this is essential to the success of lattice gauge theory.  However, we are interested in a Lorentzian world, and a Lorentzian lattice with equal time and space steps has no such enhanced symmetry.  In terms of symmetry this is better modeled by a Euclidean lattice with unequal steps (heuristically the ratio of steps is the imaginary $i$), in which case the calculation of GRP confirms the large effect seen in CPSUV.  

One might try to define the Lorentzian theory by continuation from the Euclidean lattice theory, and so benefit from the symmetry of the latter.  However, a lattice propagator has an analytic structure not consistent with unitarity.   More generally it is the requirement that a relativistic quantum theory be {\it both} Lorentz-invariant within observed limits {\it and} unitary that severely constrains the set of possible theories; if either requirement is dropped the number of possibilities multiplies enormously.  We are fortunate to have these constraints as guidance in the construction of quantum gravity.

The second model is a Lorentzian continuum theory with Pauli-Villars-like propagator
\begin{equation}
\frac{1}{-k_0^2 + \vec k^2 + m^2} - \frac{1}{-k_0^2 + \vec k^2 + \frac{m^4}{\vec k^2 + M^2} + M^2} \,. \label{pv}
\end{equation}
It is found that the induced Lorentz violation is of order $m^4/M^4$, which can be acceptably small.  However, it should be noted that the Lorentz-violating term, in the second part of the propagator, is of relative order $m^4/M^4$ at all scales, so the small breaking is put in by hand.  For the gravitational models investigated in CPSUV, the breaking is of order one at the Planck scale.  If the Lorentz-breaking term in the propagator~(\ref{pv}) is enhanced to order one at the Planck scale, then so is the induced low-energy breaking, in agreement with the results of CPSUV.  

To summarize, the question investigated by CPSUV was {``O(1) Lorentz violations at the Planck scale: do they lead to unacceptably large effects?''}  The question addressed by the models of GRP is quite different: ``Small Lorentz violations {\sl at all scales}: do they lead to unacceptably large effects?''  

To see why the first question is the relevant one, recall that one of the main issues is the possibility of observing nonstandard optical effects in high energy propagation~\cite{aemns,gp}.  In order for this to be observable and consistent with standard tests of Lorentz invariance, the effect must grow as a power of $E/M_{\rm Planck}$, but this is precisely the possibility ruled out by CPSUV: any Lorentz breaking at high energy feeds down to the Standard Model scale essentially without suppression.

Further, we note that essentially all proposed theories of Lorentz-violating quantum gravity  are of the type considered by \mbox{CPSUV}.  For example, if one formulates the theory canonically, without a Lorentz-invariant starting point, then Lorentz breaking is generically large at the Planck scale.  What CPSUV show is that one cannot then rely on dimensional analysis to guarantee consistency with observation at ordinary energies: there is a strong burden on the proponents of such a theory to demonstrate that low energy Lorentz invariance is restored to high accuracy.

We note in passing that there is one class of theories that evades the problem, namely those in which some other symmetry forbids the dimension 4 Lorentz-violating terms.  Supersymmetry is the one known example~\cite{np,jr}, in which both the space and time derivative kinetic terms arise from a single superfield.  If supersymmetry is exact at the Planck scale and broken only at a much lower scale, then the observed Lorentz breaking may be suppressed to a sufficient degree.  (Again, if the supersymmetry breaking is large at the Planck scale then the low energy Lorentz and supersymmetry breaking will be large as well).  Aside from this, the argument of CPSUV applies to all known models of observable high energy Lorentz breaking, and all known models of Lorentz-breaking quantum gravity.

GRP also give another argument to suggest that their result holds more generally, around Eq.~(7) and more fully in Sec.~IV.B of Ref.~\cite{gpr}, namely that if measurements are made with physical rods and clocks the observed Lorentz violation is small.  However, this was not applied to the relevant observations.  That is, GRP discuss the direct measurement of the Lorentz-violating propagator.  Rather, CPSUV look at the violation induced on other fields due to loops of this propagator: if one applies the same analysis to these fields then the result of CPSUV is reproduced.

 GRP attribute their result to a non-perturbative property of their models, but as we have seen it is in fact due to special features and not true in more general models that are equally `non-perturbative.'  The effective field theory argument in the first paragraph is not perturbative in nature, it is based on the Wilsonian framework that forms the basis for the non-perturbative understanding of quantum field theory.  Nor is background independence an issue.  The tests of Lorentz invariance take place in a specific nearly flat and classical background.  Any background-independent theory must reproduce the many successful tests of effective quantum field theory in this same background, and so is subject to the same constraints.

\ack
I thank Yuri Bonder, John Collins, Rodolfo Gambini, Don Marolf, Alejandro Perez, Jorge Pullin, and Edward Witten for communications and discussions.  This work was supported in part by NSF grants PHY05-51164 and PHY07-57035, and by FQXi grant RFP3-1017.

\end{document}